# Ultra-Low Delta-v Objects and the Human Exploration of Asteroids

Martin Elvis[1] and Jonathan McDowell
Harvard-Smithsonian Center for Astrophysics, 60 Garden St. Cambridge MA 02138, USA

*Jeffrey A. Hoffman and Richard P. Binzel*

Massachusetts Institute of Technology, 77 Massachusetts Avenue, Cambridge MA 02139, USA

**ABSTRACT**

Missions to near-Earth objects (NEOs) are key destinations in NASA's new "Flexible Path" approach. NEOs are also of interest for science, for the hazards they pose, and for their resources. We emphasize the importance of ultra-low delta-v from LEO to NEO rendezvous as a target selection criterion, as this choice can greatly increase the payload to the NEO. Few such ultra-low delta-v NEOs are currently known; only 65 of the 6699 known NEOs (March2010) have delta-v <4.5 km/s, 2/3 of typical LEO-NEO delta-v. Even these are small and hard to recover. Other criteria – short transit times, long launch windows, a robust abort capability, and a safe environment for proximity operations – will further limit the list of accessible objects. Potentially there are at least an order of magnitude more ultra-low delta v NEOs but, to find them all on a short enough timescale (before 2025) requires a dedicated survey in the optical or mid-IR, optimally from a Venus-like orbit because of the short synodic period for NEOs in that orbit, plus long arc determination of their orbits.

## 1. Introduction

One of the most accessible goals for human spaceflight (Augustine et al, 2009) is a rendezvous with a near-Earth object (NEO). NEOs now hold a key position in NASA's plans for human spaceflight (Obama, 2010), as a destination and a way station to exploring the inner Solar System, including Mars. Choosing the initial targets for human visits to NEOs has become a matter of immediate concern. There are some 100,000 NEOs of 100m diameter or more (Bottke et al. 2007), of which just over 6000 are now known.

NEOs are interesting for several reasons in addition to human exploration:

*1. Science*: investigating the origins of the Solar System and of life

*2. Hazards:* finding NEOs that could impact the Earth as a prelude to deflecting them.

*3. Resources:* in the long term NEOs contain the most accessible resources in space, for propellant, life support, and construction materials.

---

[1] Corresponding author: elvis@cfa.harvard.edu; phone: +1 617 495 7442; fax: +1 617 495 7356.

Each of these objectives selects a different subset of all the NEOs. For example, Mueller et al. (2011) emphasize the primitive, volatile rich, NEOs that satisfy both (1) and (3) above. All NEO selections emphasize low delta-v. Here we investigate the ultra-low delta-v tail of the NEO population, for which the energy requirements are lowest. With this constraint added to others - long launch windows, a robust abort capability, safe proximity operations environments (see sections 8, 9) – the known ultra-low delta-v population is currently very small.

## 2. Payload Gain

As is well known, the rocket equation translates a modestly lower delta-v into a significantly larger payload gain:

$$payload\ gain = e^{(dv - <dv>)/v(ex)},$$

where dv is the delta-v for a particular NEO, <dv> is the peak delta-v for the known NEO population and v(ex) is the effective rocket exhaust velocity ($I_{sp}*g$) for which we use 4.4 km/s, the value for both the RL-10 and the J-2X, the 2010-era LH2/LOX engines. This calculation neglects the mass of the inert upper stage and so gives the minimum gain.

Figure 2 shows the distribution of payload gain for missions to ultra-low velocity (<4.5 km/s, see Section 4) NEOs compared to payload at peak delta-v, for a fixed total mass in LEO, taking into account a fraction *r* of inert non-payload mass (empty rocket stages, etc) for *r* = 0 (blue), 0.1 (red). Taking the inert mass into account accentuates the advantage of the low velocity NEOs.

The importance of the few NEOs for delivering a large payload fractions to an NEO is shown in figure 3, for *r* = 0, 0.1, 0.2. A fixed total mass in LEO, and an exhaust velocity 4.4 km/s (LH2 high energy propellant) were assumed. The expressions of Shoemaker and Helin (1978) were used to calculate delta-v from the orbital elements. A more realistic treatment would treat the injection and rendezvous burns separately; the rendezvous burn would probably use a storable, lower energy propellant.

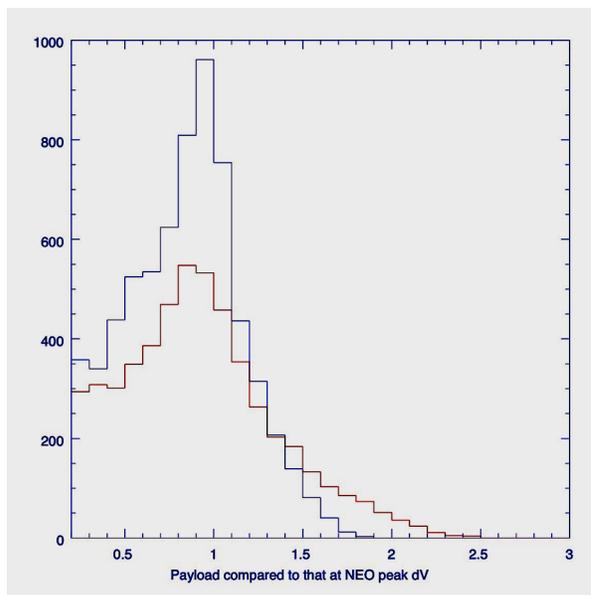

Figure 1. Payload gain for missions to low velocity NEOs compared to payload at peak delta-V, for a fixed total mass in LEO, taking into account a fraction r of inert non-payload mass (empty rocket stages, etc) for r = 0 (upper, blue, histogram), 0.1 (lower, red, histogram). Taking the inert mass into account accentuates the advantage of the low velocity NEOs.

We have compared this formula with a more careful approach, separating the two burns (at LEO and near NEO), using cryogenic fuel for the first burn and a bi-propellant for the second. We used two realistic cases for the mass fractions in the inert upper stage and engine impulses: Apollo 17, launched in 1972 with a 10% inert upper stage mass fraction, and the 2005 Mars Reconnaissance Orbiter (MRO), which used an Atlas 5 launcher, with an inert upper stage mass of 22%[2]. The gain for Apollo 17 was the factor 2.0 found from the simple treatment above, while for the larger inert upper stage mass fraction, MRO, case, the gain rose to a factor 4.4. Real mission gains will then depend on mission design specifics, but will not be less than discussed here.

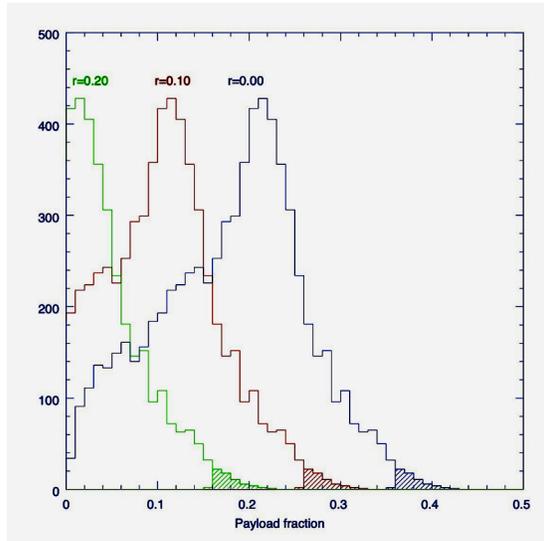

Figure 2. Histogram of mission payload fraction for all cataloged NEOs, assuming a fixed total mass in LEO, and assuming burns with exhaust velocity 4.4 km/s (LH2 high energy propellant), taking into account $r$ = 0, 0.1, 0.2. Missions to NEOs with delta-v less than 4.5 km/s are shaded, showing the significant payload advantage relative to the peak of the distribution.

### 3. Value of Payload Gain

A factor 2 gain in payload makes a major difference to the mission design, and even to the mission architecture. Crawley and Mindell (2010) discuss a system that could launch 30 mt from LEO to Earth escape; with in-orbit refueling the same launch vehicle could inject 90 mt to Earth escape. In-orbit refueling is thus a highly desirable technology. It is, however, likely to be expensive to develop and could cause budget and schedule pressures if it must be developed in parallel with new launcher systems. It is undesirable to have in-orbit re-fuelling on critical path for a human NEO mission.

---

[2] Figures for Apollo from Flight Evaluation Report for SA-512 and from Apollo 17 internal mission reports. Figures for MRO from JPL MRO Mission Design document and generic Atlas V figures from Atlas V Launch Services Users Guide, March 2010. Some are estimates, while Apollo numbers are actual reported values (except for Isp which is nominal).

Choosing an ultra-low delta-v NEO destination may enable early human NEO missions without in-orbit refueling, allowing this technology to be developed asynchronously with other crucial systems. Later missions to visit more typical 6km/s NEOs would then be enabled with in-orbit refueling of the same launch vehicle. Later destinations can be chosen for scientific, hazard mitigation, or resource purposes.

## 4. NEO delta-v Distribution

NEOs are defined to lie within Mars orbit (perihelion, q<1.3), but not wholly within Earth orbit (aphelion, Q>0.983). But an NEO on a circular 1.2AU orbit is not of interest for early human exploration, as it has high delta-v.

The delta-v of interest is the change in velocity needed to go from Low Earth orbit (LEO) to a NEO rendezvous orbit using a Hohman transfer orbit. The LEO-NEO delta-v values for known NEOs has been compiled by Benner (2010), using the Shoemaker and Helin (1978) formalism. Figure 1 shows the distribution, which has a strong peak at 6.65 km/s, with a low delta-v tail. Benner's list does not address the round trip delta-v. Detailed computer intensive simulations are needed to calculate these values accurately (Cheng 2011). In this simple listing, Benner does not allow for other important variables, such as a feasibly short round trip transit time. However, for our purposes the list provides a useful proof of concept.

Benner's list contains 6699 NEOs as of March 2010. Of these, just 65, 1%, have delta-v <4.5 km/s. We call these the 'ultra-low delta-v NEOs'.

A factor 1.5 in delta-v, from low tail to peak, has no importance to most goals for which NEOs might be visited. However for human exploration this is an important factor, because of the factor 2 gain in payload to NEO rendezvous orbit, discussed above.

Figure 3. Distribution of LEO-NEO delta-v for NEOs [3].

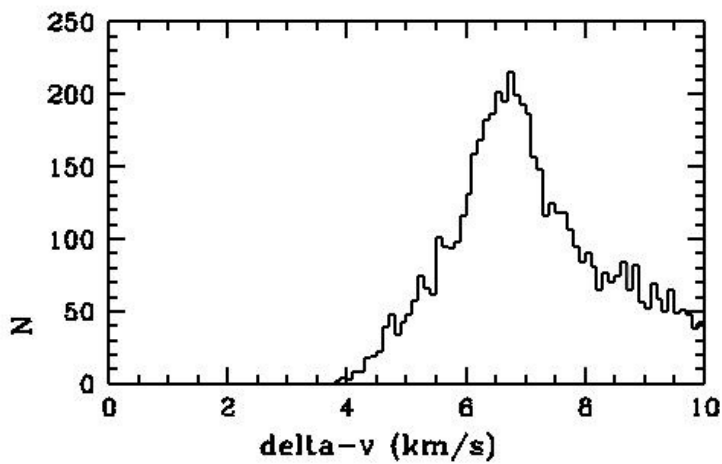

## 5. Ultra-low delta-v NEOs

The distribution of orbital semi-major axis (*a*) and eccentricity (*e*) of the 65 known ultra-low delta-v NEOs is shown in figure 4, together with other NEOs.

Two features are striking:

1. *a* and *e* are Earth-like to within a small factor. Only a small subset of all NEOs fit on this plot. Not shown is the orbital inclination (*i*) of the ultra-low delta-v NEOs, which has a mean of 2.3 degrees, with none lying above 7 degrees, while the whole NEO population has a mean of ~14.1 degrees, and a distribution ranging up to >40 degrees. Because they have such Earth-like orbits, most ultra-low delta-v NEOs drift towards or away from the Earth slowly, having synodic periods of 20 years or more.

2. Their H magnitudes range from 20 to 30, with a mean of 25.9, compared with 21.1 for all NEOs. Hence almost all known ultra-low delta-v NEOs have a nominal diameter <140m (H>22). Being so faint makes them hard to track, so that few have well-determined orbits. An example of the problem is the candidate NEO discussed for a human mission is 1999 AO10 (Farquhar et al, 2008; Abell et al, 2009).

Of this asteroid, B. Marsden (2010, private communication) said: *"[1999 AO10] was observed for only a month and spends most of its time pretty much behind the Sun as seen from the Earth. By far the best opportunity to observe it again is at the very end of 2025 and beginning of 2026, but the uncertainty in its sky position then is some degrees. And on the proposed launch date [24 Sept 2025] it would still be no brighter than mag 26 and located only 40 deg from the Sun. ... it could be found in early 2019, when it should be around mag 22.5 at around 90 deg from the Sun. If it is not found in 2019, I would not send the mission in 2025."*

Uncertainty in the position of a faint object as large as degrees makes it very hard to recover. At R=22.5 there are ~20,000 objects/sq.deg. in the sky. The large, wide-field of view LSST (see below) would be able to recover AO10, although the 2019 location puts AO10 at quadrature (i.e. along the dawn/dusk terminator line), which will make the necessary faint imaging more challenging.

Figure 4. Distribution of semi-major axis, a (AU) and eccentricity for NEOs (open black triangles) and ultra-low delta-v NEOs (filled red squares). The blue line moving right from (a=1, e=0) shows the aphelion=1 where the ultra-low delta-v NEOs are concentrated. [The green line moving left from (a=1, e=0) shows the perihelion=1 line.] The size of the symbol indicates the H magnitude of the NEO from H=20-30, smaller being fainter. All the ultra-low delta-v NEOs have inclinations, i<7 degrees.

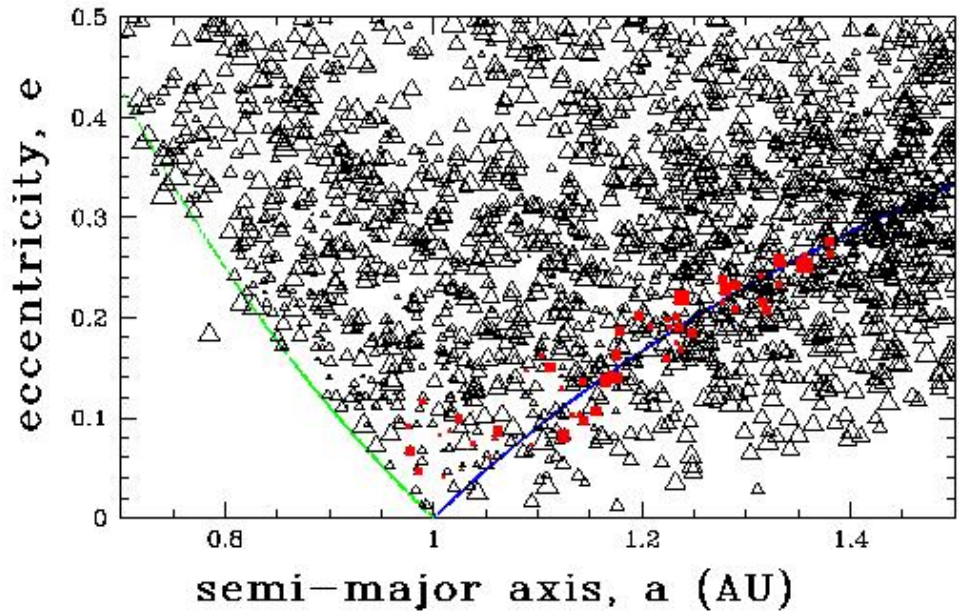

To look at the 6 most extreme low delta-v NEOs is instructive. Their known properties are given in Table.1. Almost nothing else is known about these 6 NEOs. They have no spectra taken, and no light curves measured, mostly because they are too faint. Many of them had orbits determined only roughly, as faint objects are typically tracked only briefly, most for less than 2 months, one for less than a week.

Table 1. The Six lowest delta-v NEOs known (Benner 2010)

| Name | delta-v | H | $a$ | $e$ | $i$ | Observed |
|---|---|---|---|---|---|---|
| 2006 RH120 | 3.813 | 29.5 | 1.033 | 0.025 | 0.6 | 281 |
| 2007 UN12 | 3.823 | 28.7 | 1.054 | 0.060 | 0.2 | 25 |
| 2008 HU4 | 3.927 | 28.2 | 1.093 | 0.073 | 1.3 | 41 |
| 2008 EA9 | 3.962 | 27.7 | 1.059 | 0.080 | 0.4 | 36 |
| 1991 VG | 3.998 | 28.5 | 1.027 | 0.049 | 1.4 | 173 |
| 2008 UA202 | 4.029 | 29.4 | 1.033 | 0.068 | 0.3 | 6 |

Name: provisional designation; delta-v: LEO to NEO rendezvous (km/s); H is the V band magnitude at 1AU from both Earth and Sun, at phase 0. (H=22 is 140m dia., H=27 is 14m. dia., for a typical albedo); Orbital parameters: $a$=semi-major axis (AU); $e$ = eccentricity; $i$ = inclination (deg.); Observed = number of days over which the asteroid was tracked (Marsden B., 2010, private communication).

### 6. Ultra-low delta-v NEO Numbers

Clearly a much larger pool of ultra-low delta-v NEOs, with orbits determined over long arcs, is needed in order to have a suitable list of targets for human exploration missions. There is no physical reason that larger diameter ultra-low delta-v NEOs should not exist among the uncataloged ~95% of NEOs.

However, ultra-low delta-v NEOs are not readily found. Their closely Earth-like orbits mean that most of the time they are in the daytime sky, as seen from the Earth, and so are effectively undetectable. As they approach within <1AU of the Earth they start to lie near quadrature, and so come into the dawn or dusk sky on Earth. The strong scattered sunlight background makes optical surveys toward the dawn or dusk much less sensitive and, in practice, surveys do not look in these directions, preferring to observe where the sky is dark, within 45 degrees, and at most 60 degrees, of the anti-Sun, opposition, direction. As a consequence the lowest delta-v NEOs are undercounted by current surveys, and the factor by which they are undercounted is not yet known.

Harris (2007) estimates that there are ~100,000 NEOs of 140 m diameter or larger (H<22). Of 4247 objects with H<22 from Benner (2010), there are just 2 with delta-v < 4.5 km/s. Harris (2007) predicts ~$10^7$ NEOs with H<27 (diameters 14m or larger), comparable to the 6 lowest delta-v NEOs.

The WISE spacecraft (Wright, 2008) scanned the sky around the terminator line in the mid-infrared (mid-IR) and is efficient at finding NEOs (Mainzer et al, 2010; Grav et al, 2010). By the end of the 10-month WISE mission it will be possible to estimate the ultra-low delta-v population. WISE will however only detect a few percent of the ultra-low delta-v population because of its short life.

Pan-STARRS-1 (PS1) is a ground-based optical survey using a 1.5m diameter telescope with a wide (7 sq.deg.) field of view that is surveying the sky for 2.5-3 years beginning May 2010 (Kaiser et al, 2002). One of the PS1 Key Projects is KP1 "Populations of Objects in the Inner Solar System". This survey emphasizes the discovery of NEOs. By concentrating on quadrature, called the NEO 'sweet spot' (Chesley and Spahr, 2004), KP1 expects to detect >99% of NEOs down to 300m diameter that come into range during the 3 year program. Objects with longer synodic periods, including most ultra-low delta-v NEOs, will be strongly undersampled. Nonetheless, PS1/KP1 will define the size of the ultra-low delta-v NEO population well.

### 7. Other Factors affecting human accessible NEOs

A large population of ultra-low delta-v NEOs is needed because not all of them will qualify as accessible. Other factors affecting operations, crew safety and proximity operations simplicity will reduce the final sample (Binzel et al. 2004).

*Rotation:* This is the largest factor. The surfaces of small NEOs (e.g. *25143 Itokawa*; Demura et al. 2006) can be highly irregular on both large and small scales, including boulders emerging 10s of meters (e.g. *Yoshinodai, Pencil*; Saito et al. 2006). Astronauts maneuvering within 10s of meters of the surface of a rapidly rotating asteroid would be in hazard[3]. Attachment to their surfaces is difficult given their microgravity (Wilcox, 2010).

---

[3] see full rotation of Eros from NEAR at JHU/APL: http://near.jhuapl.edu/iod/20010205/index.htm

Most NEOs will be small, as their numbers increase as roughly the inverse square of their diameters (Harris, 2007). Smaller asteroids rotate faster (Binzel et al, 1989; Harris, 2007). While above ~250m dia. asteroids are limited by their tensile strength to periods of ~2 hours or greater, about half of 100-250m dia. asteroids have shorter periods, down to a few minutes.

*Companions:* Orbiting companions to asteroids, when close, constitute an extreme case of an irregular surface. More distant companions increase the stand-off distance for the primary crew exploration vehicle and longer transit times to the NEO from the vehicle for astronauts on EVAs. Some 1/6 of NEOs are binaries down to current detection limits (Walsh and Richardson, 2008).

*Wobble:* In some cases, particularly at the smallest sizes, NEOs do not rotate about their principal moment of inertia axis (shortest axis). More specifically, the rotation axis is not aligned with a body axis. Thus the asteroid is in a state of free precession which effectively means the asteroid is "wobbling" or "tumbling" rather than being in a simple spin state. Such NEOs pose additional hazards.

*Morphology:* A more spherical asteroid poses fewer hazards to astronauts, while a highly elongated 'bone-shaped' morphology (e.g. 216 Kleopatra, [Ostro et al, 2000]), could provide useful artificial gravity if astronauts land on one of its approaching ends.

*Volatiles*: If the NEO is a dead comet, volatiles may lie close to the surface and could be exposed by human activities. Whether their sublimation would be sufficiently explosive to cause a hazard is an open question. An impactor might test for this.

## 8. Launch and Return Windows

The NEOs selected for human missions, at least at first, will require both long launch windows, and a robust abort capability, i.e. a long return window with achievable delta-v – the latter requirement has been emphasized by Farquhar et al. (2008).

With new systems launch slips are more likely, so it is prudent to select an NEO with a 3-6 month launch window for the first crewed NEO mission. Alternatively, a succession of closely spaced good targets could substitute, so long as the mission profile was sufficiently similar. For example, 1999 AO10 has a second launch window 3 months after the first, but the flight time is 30 days longer (Abell et al, 2009), which may or may not be within the mission architecture capabilities.

For crew safety a mission abort must be possible at all times during the mission. The 2025 mission to 1999 AO10 allows a return to Earth one week after the Earth escape maneuver (Farquhar et al, 2008). On the other hand, a human visit to an asteroid should allow time for the human capabilities of exploration, discovery and adaptability to be exercised. A restricted at-asteroid stay, e.g. less than 2 weeks, would strongly limit the use of human capabilities. An at-asteroid stay of a month begins to allow for true exploration. Jones et al. (2010) have noted that a larger accessible target list set helps to shorten mission duration.

In addition, Johnson (2009) emphasizes the need for a low return entry velocity (<12km/s). Abell et al. (2009) looked for NEOs accessible to the Constellation architecture between 2020 and 2035. Out of 1200 candidates they identified 12 opportunities (3 NEOs had 2). The brightest had H=23.4 (~40m dia.), highlighting the question 'should the asteroid be bigger than the spacecraft?', and recalls the difficulty of re-acquiring small NEOs noted earlier. Requiring a diameter of at least 70m (H< 23.5), Johnson (2009) finds 6 candidates.

Clearly we need a much larger NEO sample in order to have a sufficient sample of good targets.

## 9. Ultra-low delta-v NEO Specific Surveys

The choice of 2025 as a target date for NASA to have the capability to undertake a human mission to a NEO (Obama, 2010) brings a new exigency to finding a larger sample of targets.

To enable a timely and informed choice of targets, a survey for the bulk of the 100,000 NEOs with dia.>140m needs to be complete by ~2020. This implies a mean discovery rate of 10,000/year, about 10 times the current rate.

The Large Synoptic Survey Telescope (LSST) is planned to reach r(AB)=24.5 over 15,000sq.deg every 3 nights, and will find 80% of NEOs >140 m dia. in 10 years of surveying and, potentially, 90% after 12 years if 15% of the observing were optimized for this search. Uniquely the LSST high quality (5milli-mag) photometry in 6 optical (0.3-0.9micron) bands (named *u,g,r,i,z,y*) will give composition, spin state and shape estimates for the brighter NEOs (LSST, [Jones et al, 2008]). In 12 years roughly half the ultra-low delta-v NEOs will have come within range. LSST is currently planned to begin surveying in 2017, though this is contingent on obtaining funding (Ivezic et al, 2007). This is rather late for the NASA Exploration program.

As emphasized above, ground-based surveys are hampered by the dawn/dusk/daylight location of most ultra-low delta-v NEOs. Space based surveys are less limited and so are preferred.

The long synodic period of ultra-low delta-v NEOs affects survey strategy. Because the gap between the survey and the first expedition will be 5 years or more, and longer for later missions, the survey needs to span the entirety of the Earth's orbit; an ultra-low delta-v NEO that comes near the Earth in 10 years time is now behind the Sun.

This special feature of ultra-low delta-v NEOs points to a survey carried out from a Venus-like orbit (~0.7AU). Venus has a 584 day synodic period, so that employing three passes to get high survey completeness takes 4.8 years (Reitsema & Arentz 2009).

Both optical and thermal infrared surveys have been considered (e.g. NASA 2007) at sizes comparable to Kepler or Spitzer. The infrared has the advantages of providing a more model independent size estimate, and of being sensitive to low albedo asteroids.

If the first of the proposed 'Robotic Precursor Missions' were a Venus-orbit NEO survey, with selection in FY2012, a 4 year build phase and a 5 year baseline operation phase, then a catalog of ~100,000 NEOs could be ready by 2020. Estimates of the cost of such a mission are not yet certain, but seem likely to be Discovery-class, and to fit within the proposed Exploration Robotic Precursor Mission (xPRM) envelope (NASA FY2011 Budget Request).

## 10. NEO Survey Value

Each of the reasons to explore asteroids benefits from a ultra-low delta-v NEO specific survey.

*Human Exploration*: Having the largest possible choice of destinations for a human NEO mission enhances: payload, operational flexibility, safety and scientific value. By decreasing the requirements on the Earth escape launch vehicle some technologies can be removed from the critical path, increasing the probability of mission success and easing budgetary pressures by not requiring parallel, but rather serial, development.

*Hazards*: An early survey could fulfill the Congressional mandate to find 90% of 140m dia. NEOs within 15 years (George E. Brown, Jr. NEO Survey Act, Public Law No. 109-155), signed into law by President G.W. Bush on December 30, 2005.

With good orbits all asteroids will be clearly either hazardous or not, at effectively 100% confidence for the next century, or longer, solving the "potentially hazardous objects" (PHOs) question definitively. Any truly hazardous objects can then be 'tagged and towed'.

*Resources*: Such a survey will locate the most accessible space resources, a 21$^{st}$ century Lewis & Clark view of our space back yard. If the survey included a spectroscopic component the nature of these resources would become well known.

*Science*: The number of known NEOs is now somewhat over 6,000. A dedicated survey will increase the known population by more than an order of magnitude. This is similar to the factor by which the Sloan Digital Sky Survey (Gunn et al, 2006) increased the known populations of galaxies and quasars in extragalactic astrophysics. As in that case, a qualitative, revolutionary, change in NEO science will follow. Population studies will uncover the origins of families of NEOs.

## 11. Summary and Conclusions

Human exploration of NEOs requires a number of specific properties in the targets. In particular, ultra-low delta-v (LEO-NEO ~4km/s) produces payload increases by a factor 2 relative to a typical NEO. Such a gain can have important implications for mission architecture, schedule risks, and the funding profile. In a future paper we will explore the volumes of *a,e,i* parameter space for ultra low delta-v NEOs.

At present only a handful of such ultra-low delta-v NEOs are known. The complete population is however much larger. Ground-based telescopes can characterize NEOs, but a dedicated robotic precursor mission comprising a Venus-orbit optical or infrared survey seems to be needed to find all ultra-low delta-v NEOs with diameter >140m. If this were carried out by ~2020 it would enable timely target selection for the 2025 goal for a first human mission.

**Acknowledgments**

This paper is based on a talk given at the April 2010 meeting of the Dynamical Astronomy Division of the American Astronomical Society. We thank the late Brian Marsden, Alan Harris, Tim Spahr, Zlatko Ivezić and B.C. Crandall for valuable assistance and advice. This work was supported in part by NASA contract NAS8-39073.